\documentclass[fleqn,usenatbib,useAMS]{mnras}

\usepackage{graphicx}	
\usepackage{amsmath}	
\usepackage{amssymb}	
\usepackage{multicol}        
\usepackage{bm}		
\usepackage{pdflscape}	
\usepackage{csvsimple}

\usepackage[T1]{fontenc}
\usepackage{ae,aecompl}

\usepackage{newtxtext,newtxmath}

\title[The Galactic population of energetic pulsars]
{The Galactic population and properties of young, highly-energetic pulsars}
\author[Johnston et al.]
{Simon Johnston$^{1}$\thanks{E-mail: simon.johnston@csiro.au},
D. A. Smith$^2$,
A. Karastergiou$^{3,4}$,
M. Kramer$^5$
\\
$^{1}$CSIRO Astronomy and Space Science, Australia Telescope National Facility, PO~Box~76, Epping NSW~1710, Australia\\
$^2$Centre d'\'Etudes Nucl\'eaires de Bordeaux Gradignan, IN2P3/CNRS, Universit\'e Bordeaux, 33175 Gradignan, France\\
$^{3}$Department of Astrophysics, University of Oxford, Denys Wilkinson Building, Keble Road, Oxford OX1 3RH, UK\\
$^{4}$Department of Physics and Electronics, Rhodes University, PO Box 94, Grahamstown 6140, South Africa\\
$^{5}$Max-Planck-Institut f\"ur Radioastronomie (MPIfR), Auf dem H\"ugel 69, D-53121 Bonn, Germany\\
}
\date{Last updated; in original form}

\pubyear{2020}

\begin{document}
\label{firstpage}
\pagerange{\pageref{firstpage}--\pageref{lastpage}}
\maketitle

\begin{abstract}
The population of young, non-recycled pulsars with spin down energies $\dot{E} >10^{35}$~erg s$^{-1}$ is sampled predominantly at $\gamma$-ray and radio wavelengths. A total of 137 such pulsars are known, with partial overlap between the sources detectable in radio and $\gamma$-rays. We use a very small set of assumptions in an attempt to test whether the observed pulsar sample can be explained by a single underlying population of neutron stars. For radio emission we assume a canonical conal beam with a fixed emission height of 300~km across all spin periods and a luminosity law which depends on $\dot{E}^{0.25}$. For $\gamma$-ray emission we assume the outer-gap model and a luminosity law which depends on $\dot{E}^{0.5}$. We synthesise a population of fast-spinning pulsars with a birth rate of one per 100 years. We find that this simple model can reproduce most characteristics of the observed population with two caveats. The first is a deficit of $\gamma$-ray pulsars at the highest $\dot{E}$ which we surmise to be an observational selection effect due to the difficulties of finding $\gamma$-ray pulsars in the presence of glitches without prior knowledge from radio frequencies. The second is a deficit of radio pulsars with interpulse emission, which may be related to radio emission physics. We discuss the implications of these findings.
\end{abstract}



\section{Introduction}
Pulsars are fast-spinning neutron stars which emit radiation across the electromagnetic spectrum. As a result they lose energy and their rotation rate decreases. For a pulsar with spin period $P$ and spin-down rate $\dot{P}$, its rate of energy loss, $\dot{E}$, is given by
\begin{equation}
    \dot{E} = 4\,\,\pi^2\,\,I\,\,\frac{\dot{P}}{P^3}
\end{equation}
where $I$ is the moment of inertia. Although the equation of state of nuclear matter is not well known and a range of masses and radii amongst the pulsar population is likely, we expect the moment-of-inertia to be close to a canonical value of $I=10^{45}$~gcm$^2$, consistent with experimental constraints \citep[e.g.][]{eosligo}. 

The youngest, most energetic pulsars have $\dot{E}\sim 10^{38.5}$~erg s$^{-1}$ and, in radio pulsars, the smallest observed values of $\dot{E}$ are some $10^9$ times lower. In $\gamma$-rays, very few non-recycled pulsars are known with  $\dot{E}<10^{34}$~erg s$^{-1}$.
In contrast, the energy output at radio frequencies is insignificant compared to that in the $\gamma$-ray band. Yet, until about 10 years ago, most of what was known about the population of rotation-powered neutron stars was known from radio surveys. The advent of the Large Area Telescope (LAT) on the {\it Fermi} space observatory has opened another window, revealing not only an increasing sample of neutron stars that can only be detected via their high energy emission, but that the different emission process with its different beaming characteristics also provides an independent view onto the same underlying population. By combining geometrical information and constraints from both extremes of the electromagnetic spectrum, one can expect to derive a more complete picture and, eventually, understanding that helps to answer fundamental questions about the initial birth properties and the evolution from the young into the older bulk population of neutron stars.

Hence, in this work, we concentrate on the young, high $\dot{E}$ pulsars.
The scientific drivers for examining their properties are several-fold. They tell us about the birth rate of neutron stars in the Galaxy which can be compared with the rate of their putative progenitors, core-collapse supernovae \citep{dhk+06,kk08}. They inform us about the initial spin-period of pulsars, a topic of much contention \citep{vml+04,fk06,lsf+06,jk17}. $\dot{E}\sim10^{35}$~erg s$^{-1}$ marks a transition between $\gamma$-ray and non $\gamma$-ray pulsars with implications for magnetospheric physics \citep{mh04,wrwj09,pghg12,ps18,khkw19}. Finally, the Crab pulsar, the most energetic pulsar in the Galaxy may emit radiation at up to 100~TeV and beyond \citep{mc20} and continues to intrigue at lower energies \citep{ygl+18,bc19}.

Prior to 2008, at most 10 $\gamma$-ray pulsars were known. The launch and successful operation of the LAT on {\it Fermi} \citep{fermi} has changed this picture completely \citep[][and references therein]{4FGL} and today, 12 years after launch, more than 250 such pulsars are catalogued\footnote{https://confluence.slac.stanford.edu/display/GLAMCOG/Public+List+of+LAT-Detected+Gamma-Ray+Pulsars} \citep{2pc,sbc+19}. In conjunction with the deep radio surveys of the Galactic plane, which have yielded more than 1000 pulsars, a statistical undertaking of the population is now a viable proposition. Previous work along these lines which combined radio and $\gamma$-ray data include the papers of \citet{rmh10}, \citet{twc11}, \citet{wr11} and \citet{pghg12}, all written in the very early days of the {\it Fermi} mission. \citet{wr11} showed that fast initial spin periods and a birth rate of 1 per 60~yr were required to produce the $\gamma$-ray population. They predicted that after 10 years of {\it Fermi} operations some 120 young $\gamma$-ray pulsars would be known of which about one half would be radio quiet. \citet{rmh10} showed that there must be considerable overlap between the radio beams and the $\gamma$-ray beams and postulated high radio emission heights. \citet{pghg12} carried out a full-blown population analysis including a detailed simulation of the radio and $\gamma$-ray pulse profiles. Although they were able to reproduce many of the observational results, they under-predicted the number of $\gamma$-ray pulsars seen at the highest energies.

In this paper we consider young (non-recycled) pulsars with $\dot{E}>10^{35}$~erg s$^{-1}$. We draw the line here for three specific reasons. The first is that the detectability of $\gamma$-ray pulsars drops dramatically below this value, both because the conditions in the magnetosphere are no longer ripe for the emission of $\gamma$-rays and because the luminosity becomes low. This means a high incompleteness fraction in the observed population and makes modelling difficult. Secondly there appears to be a change in properties amongst radio pulsars at or near $\dot{E}=10^{35}$~erg s$^{-1}$. In particular, above this value, their polarization fraction is very high \citep{avh00,wj08b,jk18}, their pulse profiles follow a particular shape \citep{jw06,kj07} and the emission height may also be high \citep{jw06,kj07,rmh10,lgw+13,rwj15b}. The third reason is practical: it is much simpler to simulate the Galactic population of these pulsars which are typically less than $10^5$ years old. In particular, contentious issues such as pulsar velocities \citep{acc02,hllk05}, the possibility of magnetic field decay \citep{fk06,ycbb10,gmv14}, changes in geometry over time \citep{tm98,wj08a,jk17}, and the radio `death-line' \citep{ymj99,ztzw17} can safely be ignored.

In order to determine the underlying Galactic population of these pulsars from which the observational sample is drawn, we need to take into account three main factors. The first is the {\em beaming fraction} which tells us how much sky is illuminated by the pulsars, and hence the fraction of the population that is {\em potentially} detectable from Earth. The beaming fractions of radio and $\gamma$-ray pulsars are different and depend on the pulsar's spin parameters, geometry and location of the emission regions. The second factor is the {\em luminosity} of the pulsars with the third factor being the {\em sensitivity} of a given radio or $\gamma$-ray survey. The combination of the latter two factors tell us how many pulsars are {\em actually} detectable. We will deal with these factors in the subsequent sections. Our overall aim is to determine whether the canonical picture of a spinning pulsar suffices to reproduce the observational results without the need for an overly complex, multi-parameter, multi-assumption fit.

In Section~2 we give our observational selection, in Section~3 we outline a prescription for the beaming fraction of radio and $\gamma$-ray pulsars. Section~4 deals with the detectability of the pulsars given the surveys at both wavebands. In Section~5 we show the results from our simulations and discuss their implications in Section~6.

\section{Observational selection}
Since we concentrate on the young, Galactic pulsars, we have excluded the two energetic pulsars in the Magellanic Clouds, PSRs~B0540--69 and J0537--6910. We exclude the recycled, millisecond pulsars, many of which are also seen in $\gamma$-rays. We further exclude 6 high $\dot{E}$ pulsars discovered in the X-ray band which have no radio or $\gamma$-ray counterparts.

With these exclusions, version 1.62 of the on-line pulsar catalogue \footnote{http://www.atnf.csiro.au/people/pulsar/psrcat/}, contains 137 pulsars with $\dot{E}>10^{35.0}$~erg s$^{-1}$. Of this total, 106 are seen as radio pulsars and 95 as $\gamma$-ray pulsars. Table~\ref{tab_edot} shows the number of pulsars per $\dot{E}$ decade subdivided into radio-only ($N_r$), $\gamma$-ray only ($N_g$) and both radio and $\gamma$-ray ($N_{gr}$) along with the mean spin period of the pulsars in that decade. Several things can be noted. First, that only a single $\gamma$-ray pulsar exists at $\dot{E}>10^{37}$~erg s$^{-1}$ without a radio counterpart.  Secondly that the number of $\gamma$-ray pulsars exceeds the number of radio pulsars in the range $10^{36} < \dot{E} < 10^{37}$~erg s$^{-1}$.

Over the past 50 years, radio pulsars have been detected with a wide variety of techniques, telescopes and observing frequencies. However, for this selection of pulsars, virtually all are at low Galactic latitudes. For the modelling described below we therefore parameterise the surveys carried out by the Parkes radio telescope in the Southern Galactic plane \citep{kbm+03,lsf+06,ccb+20} and the Arecibo telescope in the northern plane \citep{cfl+06}, all of which were carried out at an observing frequency of 1.4~GHz. In addition to blind surveys of the Galactic plane, radio pulsars have also been discovered by performing very deep searches on targets of interest such as known X-ray or $\gamma$-ray pulsars and supernova remnants. This has resulted in the discovery of extremely faint radio pulsars (e.g. \citealt{crr+09}), well below the sensitivity of large-scale surveys. We note that of the 106 radio pulsars with $\dot{E}>10^{35}$~erg s$^{-1}$ that 83 were detected in these surveys with a further 12 being the result of deep radio searches of high-energy counterparts and/or supernova remnants. Of the 11 remaining, 6 are outside the survey regions and the remaining 5 below the nominal sensitivity threshold. We return to this in Section~4.2.

The discovery of young $\gamma$-ray pulsars by the {\it Fermi} LAT follows two different routes. First, the ephemerides from known pulsars are used to fold the $\gamma$-ray photons (e.g. \citealt{sgc+08,sbc+19}). Secondly, blind searches are made in the photon data directly (e.g. \citealt{pga+12,cwp+17}). These two methods have different sensitivity thresholds; it is significantly easier to find pulsars already known at other wavelengths. We return to this in Section~4.3.

\section{Beaming fraction}
\subsection{Radio pulsars}
In the prevailing (observational) model of radio pulsars, the radio emission arises from near the pulsar surface across the bundle of open field lines \citep{lk05}. This yields a simple geometrical relationship between the half opening angle of the emission cone ($\rho$), the emission height ($h_{em}$) and the spin period viz:
\begin{equation}
\rho = 3\,\,\, \sqrt{\frac{\pi \,\,\, h_{\rm em}}{2\,\,\, P\,\,\, c}}
\label{eqn:rho}
\end{equation}
with $c$ the speed of light. If the magnetic axis is inclined with respect to the rotational axis by an angle $\alpha$ then the emission cone sweeps out an area of sky which is typically much less than $4\pi$~sr. Knowledge is therefore required of the underlying $P$, $\alpha$ and $h_{em}$ distributions and their possible interdependence before the beaming fraction can be computed. Leaving $h_{em}$ to one side for the moment, we make the following assumptions. Firstly, that the observed period distribution reflects the underlying period distribution: in modern surveys there are little or no selection effects in detecting pulsar with periods in excess of 20~ms. Note that the radio and $\gamma$-ray pulsars partly overlap, and that we assume that they are different views onto the same underlying neutron star population. Hence, the period distribution for radio and $\gamma$-ray pulsars should be the same. This is also the case for the distribution for the magnetic inclination angle. Secondly, we assume in the absence of information, that $\alpha$ is randomly distributed in these objects (i.e. the probably density function follows cos$^{2}\alpha$ in the observed sample). 
\begin{table}
\caption{Number of known pulsars above $\dot{E}=10^{35}$~erg s$^{-1}$ per decade}
\label{tab_edot}
\begin{center}
    \begin{tabular}{crrrrr}
    \hline
    \hline
log$\dot{E}$ & total & $N_{r}$ & $N_g$ & $N_{gr}$ & $<P>$ \\
(erg s$^{-1}$) & & & & & (ms)\\
\hline
$>38$ & 1 & 0 & 0 & 1 & 33 \\
$37-38$ & 13 & 4 & 1 & 8 & 77 \\
$36-37$ & 45 & 8 & 12 & 25 & 118 \\
$35-36$ & 78 & 30 & 18 & 30 & 174 \\
\hline
total & 137 & 42 & 31 & 64 \\
\hline
    \end{tabular}
\end{center}
\end{table}

Much work has been carried out on determining pulsar emission heights (e.g.~ \citealt{bcw91,mr02,ml04,jk19}). Our prevailing understanding is that $h_{em}$ does not depend on the spin period, but is roughly constant with a mean value of $\sim$300~km across the population. This can be inferred from observations via Equation~\ref{eqn:rho}, as a number of studies found consistently $\rho \propto P^{-0.5}$ (e.g. \citealt{kwj+94,gl98,mg11,sbm+18}). 
We note that there is good observational evidence that emission height is largely independent of observing frequency above about 400~MHz \citep{tho91}.
Armed with the height, the period distribution and a random distribution of $\alpha$ the radio beaming fraction can therefore be computed for an ensemble of objects.

Equation~\ref{eqn:rho} describes the opening angle of a "fully-filled" open-fieldline region. In this case the width of the radio profile, $w_r$, can be computed from knowledge of $\rho$ and the geometry via:
\begin{equation}
{\rm cos}\rho = {\rm cos}\alpha\,\, {\rm cos}\zeta\,\, +\,\, {\rm sin}\alpha\,\, {\rm sin}\zeta\,\, {\rm cos}(w_r/2)
\label{eqn:wr}
\end{equation}
where $\zeta$ is the angle between the rotation axis and the line of sight in the plane of the magnetic axis.
It may not be the case that the beam is filled, and indeed, the distribution of radio emission within the emission cone is a vexed problem (e.g.~\citealt{ran93,lm88,kj07,dr15}). However, it is not overly critical for our considerations. As  long as some fraction of the beam is illuminated, the pulsar can still be detected. Here, the latitudinal extent of the actual beam is more important than the longitudinal extent (see e.g. \citealt{dkl+19}). The latter will become important, however, when we try and compare the observed widths of the radio pulsar population with predicted widths.

\subsection{$\gamma$-ray pulsars}
The three major classes for the production of $\gamma$-rays in pulsars are the outer gap (OG), two-pole caustic (TPC) and force-free (FF) models. In the OG model, $\gamma$-rays originate in a vacuum gap above the null charge surface and outwards to the light cylinder \citep{chr86,rom96}. In the TPC model, emission can extend downwards almost to the pulsar surface, with the resultant implication that the low-altitude emission is seen from one pole while the high-altitude emission is seen from the other \citep{dr03,mh04}. In the FF model, $\gamma$-rays originate in an equatorial current sheet close-to and beyond the light cylinder radius \citep{ps18,khkw19}.

In this paper we use the OG model primarily because it is a geometrical model, which makes it amenable to analytic evaluation rather than having to rely on the numerical simulations inherent in the FF model. A critical parameter in the OG model is the width of the gap, and its dependence on $\dot{E}$. We adopt the approach taken by \citet{wrwj09}, and define the width ($w_g$) of the gap as
\begin{equation}
w_g = \sqrt{\frac{10^{33}}{\dot{E}}}
\label{eqn:thick}
\end{equation}
and so as $\dot{E}$ decreases, $w$ increases. As a result the null charge line migrates upwards and the $\gamma$-ray emission sweeps out less of the sky. In OG models, therefore, emission is produced beyond a minimum emission angle $\epsilon_{og}$ from the rotation axis, where $\epsilon_{og}$ is given by
\begin{equation}
\epsilon_{og} = (75 + 100w_g) - (60 + 1/w_g)(\alpha/90)^{2(1-w_g)}
\label{eqn:og}
\end{equation}
and is absent otherwise. We note that for large values of $\dot{E}$ where equation~\ref{eqn:og} can return $\epsilon_{og}<0\degr$ we set $\epsilon_{og}=0\degr$. For the origin of these equations see \citet{wrwj09}.

\begin{figure}
\centering
\includegraphics[width=0.5\textwidth]{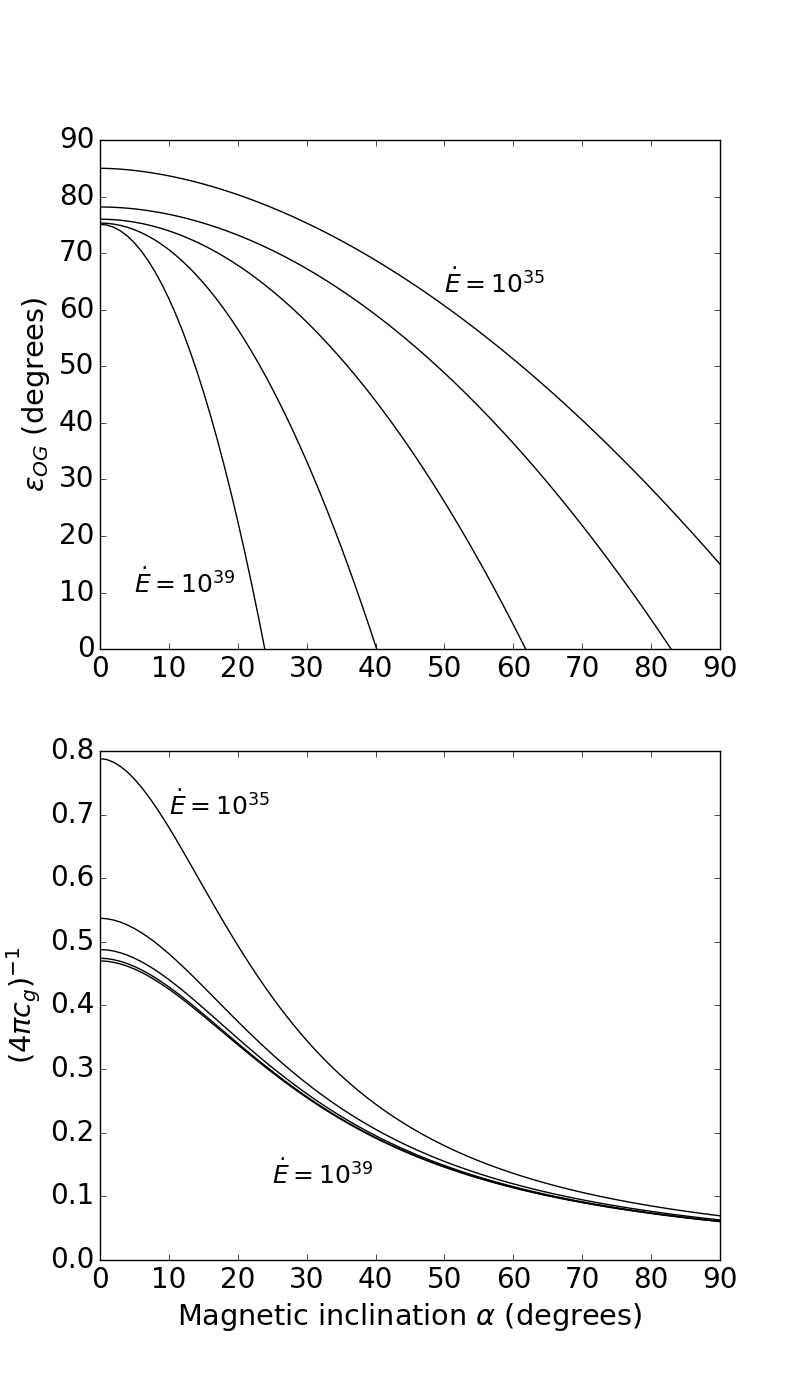}
\caption{Top frame: Minimum viewing angle measured from the rotation axis to the line-of-sight, $\epsilon_{og}$, as a function of the magnetic inclination $\alpha$ (Equation~\ref{eqn:og}). The 5 curves are for $\dot{E}$ from $10^{35}$ to $10^{39}$~erg s$^{-1}$. $\gamma$-ray emission is detectable from Earth for $(\alpha, \, \zeta)$ pairs above and to the right of the curves. Bottom frame: The factor $1/4\pi c_g$ as a function of $\alpha$ (Equations~\ref{eqn:fg1} and \ref{eqn:fg2}) for the same $\dot{E}$ values as the top panel.}
\label{fig:CgZogPlot}
\end{figure}

Unlike with the radio case, the $\gamma$-ray beaming fraction does not depend on the spin period but does depend on $\dot{E}$ which enters through equation~\ref{eqn:thick}. The top panel of Figure~\ref{fig:CgZogPlot} shows $\alpha$ versus $\epsilon_{og}$ for a range of $\dot{E}$ values. For a given $\dot{E}$, $\gamma$-ray emission is beamed towards Earth for values of ($\alpha$,$\epsilon_{og}$) above and to the right of the curves shown. Therefore when $\alpha$ is small, $\gamma$-ray emission can only be seen when the observer views close to the spin equator (i.e. $\epsilon_{og}\sim90\degr$). Conversely, for orthogonal rotators ($\alpha\sim90$\degr), the $\gamma$-ray emission covers virtually the whole sky for the entire $\dot{E}$ range.

\section{Detectability}
\subsection{Galactic distribution}
We are dealing here with only the youngest fraction of the total pulsar population. We simply assume that pulsars are born in the Galactic plane and do not move significantly over the first $\sim$100~kyr of their lifetime. This obviates the need to include the pulsar velocity distribution which is not well known \citep{acc02,hllk05}. For the Galactic radial distribution of pulsars we assume the form of the radial distribution given by \citet{lor04}
\begin{equation}
\rho_r(R) = K_r \,\,\,  R^i \,\,\, e^{-R/\sigma_r}
\label{eqn:radial}
\end{equation}
where $\rho_r(R)$ is the density of pulsars (per kpc$^2$) at radius $R$ (in kpc) from the Galactic Centre and $K_r$, $i$ and $\sigma_r$ are constants with values of 64.6~kpc$^{-2}$, 2.35 and 1.258~kpc respectively.

\subsection{Radio pulsars}
An accepted generalized form for the radio luminosity law is
\begin{equation}
L_r = L_0 \,\,\, P^{\epsilon_1} \,\,\, \dot{P}^{\epsilon_2} \,\,\,L_j
\label{lumin}
\end{equation}
where the parameters can be estimated from the observed population (e.g. \citealt{fk06,lsf+06}). Although $P$ varies by two orders of magnitude in the normal pulsar population and $\dot{P}$ varies by six orders of magnitude, $L_r$ only varies by about two orders of magnitude across the entire population. This implies already that the values of $\epsilon_1$ and $\epsilon_2$ cannot be too large. In the most recent study of the population as whole, \citet{jk17} found that $\epsilon_1 = -0.75$, $\epsilon_2 = 0.25$ gave the best fit to the data, noting that this implies $L_r \propto \dot{E}^{1/4}$. The $L_j$ term is introduced to provide scatter in $L_r$ for a given $P$ and $\dot{P}$ as seen in the observed population.

As we have $L_r \propto \dot{E}^{1/4}$ then we can then write the radio flux density at 1.4~GHz, $F_r$, of a pulsar in mJy as
\begin{equation}
F_r = \frac{9.0}{d^2}\,\,\,\frac{\dot{E}^{1/4}}{10^9}\,\,\,10^{F_j}
\label{eq:fluxden}
\end{equation}
where $d$ is the distance in kpc and $F_j$ is the scatter term which is modelled as a Gaussian with a mean of 0.0 and $\sigma=0.2$. Then, for example, a pulsar with $\dot{E}=10^{36}$~erg s$^{-1}$ at a distance of 1~kpc has a mean flux density of 9~mJy.

We can then convert $F_r$ into a detection signal-to-noise ratio (S/N) for a given pulsar survey:
\begin{equation}
{\rm S/N} = \frac{F_r}{S_0} \sqrt{\frac{P-w_r}{w_r}}
\label{snr}
\end{equation}
where the scaling factor $S_0$ reflects the survey parameters. The term inside the square root gives the advantage of detecting narrow pulse profiles (of width $w_r$) using the Fourier technique employed in pulsar surveys. Equation~\ref{eqn:wr} can be used to determine $w_r$. Detections are made when S/N~$>10$. 
The Parkes and Arecibo surveys described in Section~2 have a sensitivity of $\sim$0.15~mJy to long period pulsars in the Galactic plane and so $S_0\sim0.05$. For the deep follow-up surveys at 1.4~GHz of $\gamma$-ray sources, we assume a radio detection threshold 3 times lower.

\subsection{$\gamma$-ray pulsars}
The $\gamma$-ray luminosity for pulsars with $\dot{E}>10^{33}$~erg s$^{-1}$ is given by
\begin{equation}
L_g = \dot{E} \,\, \sqrt{\frac{10^{33}}{\dot{E}}}
\end{equation}
This luminosity law ensures that the efficiency in $\gamma$-rays drops proportionally to $\sqrt{\dot{E}}$ as expected from theory, and consistent with observations \citep{2pc,sbc+19}.

The line of sight cut of an Earth-based observer is defined by the angles $\alpha$ and $\zeta$. For a given line of sight, we first check whether $\gamma$-ray emission is beaming towards us which occurs when $\zeta>\epsilon_{og}$ (see equation~\ref{eqn:og}). If so, we define the flux correction factor, $c_g$, following \citet{wrwj09}.
\begin{equation}
c_g = 0.17 - 0.69w_g + (1.15-1.05w_g)(\alpha/90)^{1.9}
\label{eqn:fg1}
\end{equation}
The $\gamma$-ray flux as detected on Earth, $F_g$, is then simply $L_g$ corrected for the line-of-sight cut and divided by the square of the distance to a given pulsar.
\begin{equation}
F_g = \frac{1}{4\pi c_g} \,\, \frac{\dot{E}}{d^2} \sqrt{\frac{10^{33}}{\dot{E}}}
\label{eqn:fg2}
\end{equation}
The bottom panel of Figure~\ref{fig:CgZogPlot} shows $1/(4\pi c_g)$ as a function of $\alpha$ and $\dot{E}$. The correction factor is close to 0.1 for $\alpha>60\degr$, but for low $\alpha$ where the beam is confined to the rotational equator the correction factor can be more than a factor of 5 larger.

We assume that the sensitivity to pulsars at Galactic latitudes $<2$\degr\ is $4\times10^{-12}$~erg cm$^{-2}$s$^{-1}$ for known pulsars and $16\times10^{-12}$~erg cm$^{-2}$s$^{-1}$ for blind searches, scaled from \citet{2pc} to allow for increased time span. These thresholds are averages that match well both the pulsar fluxes published in \citet{4FGL} and the fact that a group of pulsars lie below the DC sensitivity limit of 4FGL. Finally then, a $\gamma$-ray pulsar is detected by the simulation if $F_g$ exceeds these values.

We note that we require that the $\gamma$-ray profile have structure and is not just a DC term adding to the background noise. Examination of the atlas shows that when $\alpha<35$\degr\ and $\zeta<85$\degr\ no pulsations are seen, and similarly for $\zeta<25$\degr\ at any $\alpha$. We therefore remove these sources from the simulation even if their $F_g$ exceeds the threshold.

\section{Simulation}
The existence of a population of radio and $\gamma$-ray pulsars at high $\dot{E}$ already informs us that there is a need for fast initial spin periods in the pulsar population (see also \citealt{wr11}). If we take a pulsar born with a spin-period of 50~ms and a magnetic field of $4.5\times10^{12}$~G then we can compute its trajectory through $P$, $\dot{P}$ (and hence $\dot{E}$) space. The spin-down of a pulsar is generally written in the form
\begin{equation}
\dot{\nu} = -K \nu^n
\label{nu}
\end{equation}
where $\nu$ and $\dot{\nu}$ are the spin frequency and its derivative, $K$ is constant and $n$ is the braking index which here we set to the canonical $n=3$ value that results if the torque that slows the neutron star is entirely due to dipole radiation (e.g. \citealt{lk05}). Under such a spin-down law, the pulsar takes 68.9~kyr to fall below an $\dot{E}$ of $10^{35}$~erg s$^{-1}$, at which point it has a spin-period of 297~ms. In this simple example therefore, with a birth rate of one such pulsar per 100~yr, there would be 689 such pulsars in the Galaxy, only 3 of which would have $\dot{E}>10^{38}$~erg s$^{-1}$.
\begin{table}
\caption{Number of pulsars in the Galaxy above $\dot{E}=10^{35}$~erg s$^{-1}$ and their mean period $P$ (in ms) under the assumptions given in Section 5.}
\label{tab_sim}
\begin{center}
    \begin{tabular}{crr}
\hline
\hline
log$\dot{E}$ & Number & $<P>$\\
(erg s$^{-1}$) &  & (ms)\\

\hline
$>38$   & 3.6   & 54 \\
$37-38$ & 40.4  & 73 \\
$36-37$ & 186.8 & 117 \\
$35-36$ & 614.5 & 197 \\
\hline
total & 845.2 \\
\hline
    \end{tabular}
\end{center}    
\end{table}

We have explored various different scenarios for the initial spin-period and magnetic field by assuming mean values of 50~ms and  $4.5\times10^{12}$~G and using Gaussian distributions with a variety of widths, $\sigma_p$ and $\sigma_b$. For the initial periods, we adopt a $\sigma_p = 10$~ms, with periods truncated at a minimum 10~ms. For the magnetic field, we use a Gaussian in log-space with $\sigma_b = 0.3$. These distributions are similar to the ones used by both \citet{wr11} and \citet{pghg12}.
Table~\ref{tab_sim} gives the results for a birth rate of 1 per 100~yr after performing $10^4$ trials; the numbers scale with the reciprocal of the birth rate. The mean spin periods from the simulation compare extremely well with those given in Table~\ref{tab_edot} for the observed population. We note that we remain agnostic to the presence or not of pulsars `injected' into the population with spin periods of 300~ms and lower magnetic fields \citep{vml+04,fk06}. These pulsars do not meet the $\dot{E}>10^{35}$~erg s$^{-1}$ threshold.
\begin{table*}
\caption{Beaming and detection fraction of radio and $\gamma$-ray pulsars above $\dot{E}=10^{35}$~erg s$^{-1}$ per decade along with the total number of detections as given by the simulation for an input birth rate of 1 pulsar per 100~years.}
\label{tab_bf2}
\begin{center}
    \begin{tabular}{crcccccccrrrr}
    \hline\hline
log($\dot{E})$ & $S_p$ & $f_r$ & $f_g$ & $f_{rg}$ & $d_r$ & $d_g$ & $t_r$ & $t_g$ & $S_t$ & $S_r$ & $S_g$ & $S_{gr}$\\
erg s$^{-1}$ & $\times10^4$ &&&&&&&&$\times10^4$  & $\times10^4$ & $\times10^4$ & $\times10^4$\\
\hline
$>38$    & 3.6 & 0.56 & 0.92 & 0.50 & 0.82 & 0.62 & 0.46 & 0.58   &  2.6  & 0.2 & 0.9 & 1.5\\
$37-38$ & 40.4 & 0.50 & 0.82 & 0.39 & 0.64 & 0.48 & 0.32 & 0.39  &  19.5 & 1.9 & 6.6 & 11.0\\
$36-37$ & 186.7 & 0.42 & 0.67 & 0.28 & 0.42 & 0.28 & 0.17 & 0.18 &  44.7 & 9.6 & 12.1 & 23.0\\
$35-36$ & 614.5 & 0.33 & 0.50 & 0.19 & 0.29 & 0.13 & 0.09 & 0.06 &  70.6 & 30.6 & 12.8 & 27.2\\
\hline
total & 845.2 && &&&&&& 137.4 & 42.3 & 32.4 & 62.7\\
\hline
    \end{tabular}
    \end{center}
\end{table*}

Having generated $845\times10^4$ pulsars ($10^4$ times the true number to avoid small number statistics) each with $P$ and $\dot{E}$ from the prescription described above, we place them in the Galaxy using the Galactic radial distribution given by equation~\ref{eqn:radial}. We then assign random $\alpha$ and $\zeta$ and compute the beaming fraction of the pulsars using the prescriptions in Sections~3.1 and 3.2. Finally, for the pulsars beaming towards us, we can compute the flux in the radio and $\gamma$-ray bands and then `detect' the pulsars above the relevant thresholds using the heuristics in Sections~4.2 and 4.3. We can then determine the number of radio-only, $\gamma$-ray only or joint radio/$\gamma$ detected pulsars.

Table~\ref{tab_bf2} gives the results. Column 1 lists the $\dot{E}$, column 2 gives the number of pulsars generated in the simulation from Table~\ref{tab_sim}. Columns 3 to 5 give the beaming fraction for radio and $\gamma$-ray pulsars and the overlap between them. Columns 6 and 7 give the detection fraction of the beamed pulsars with columns 8 and 9 then giving the total fraction of pulsars detected ($t_r$ and $t_g$), obtained by multiplying the beaming fraction by the detection fraction. The final 4 columns list the total number of pulsars detected in the simulation, ($S_t$), and the breakdown between radio, $\gamma$-ray and joint radio/$\gamma$-ray pulsars. These values depend on the input birth rate of 1 per 100 years. They can be scaled with birth rate and then compared directly with the observed values from Table~\ref{tab_edot}. We find the best fit to the observations, subject to the caveats discussed below, arises for a birth rate of 1 per $95\pm10$ years. We get the same value for {\bf both} radio and $\gamma$-ray pulsars even though their luminosity equations and their detectability are completely independent. Figure~\ref{fig:summary} shows the results in graphical form.

\section{Discussion}
Table~\ref{tab_bf2} shows that the beaming fraction in $\gamma$-rays is very high at high $\dot{E}$ but drops quickly whereas the radio beaming fraction evolves more slowly with $\dot{E}$. This is because in the radio, $f_r$ depends only on $P^{-0.5}$ which varies only slowly with $\dot{E}$ whereas $f_g$ depends directly on $\dot{E}^{0.5}$. The overlap between the radio and $\gamma$-ray beams ($f_{rg}$) falls by almost a factor of 3 from $\dot{E}=10^{38}$ to $\dot{E}=10^{35}$~erg s$^{-1}$. Table~\ref{tab_bf2} also shows that the detected fraction of radio and $\gamma$-ray pulsars is similar for $\dot{E}>10^{36}$~erg s$^{-1}$; below this value the radio surveys win out. As a result then, one expects to detect more $\gamma$-ray pulsars than radio pulsars for $\dot{E}>10^{37}$~erg s$^{-1}$ whereas there should be more radio than $\gamma$-ray pulsars for $\dot{E}<10^{36}$~erg s$^{-1}$. 

\begin{figure}
\centering
\includegraphics[width=0.52\textwidth]{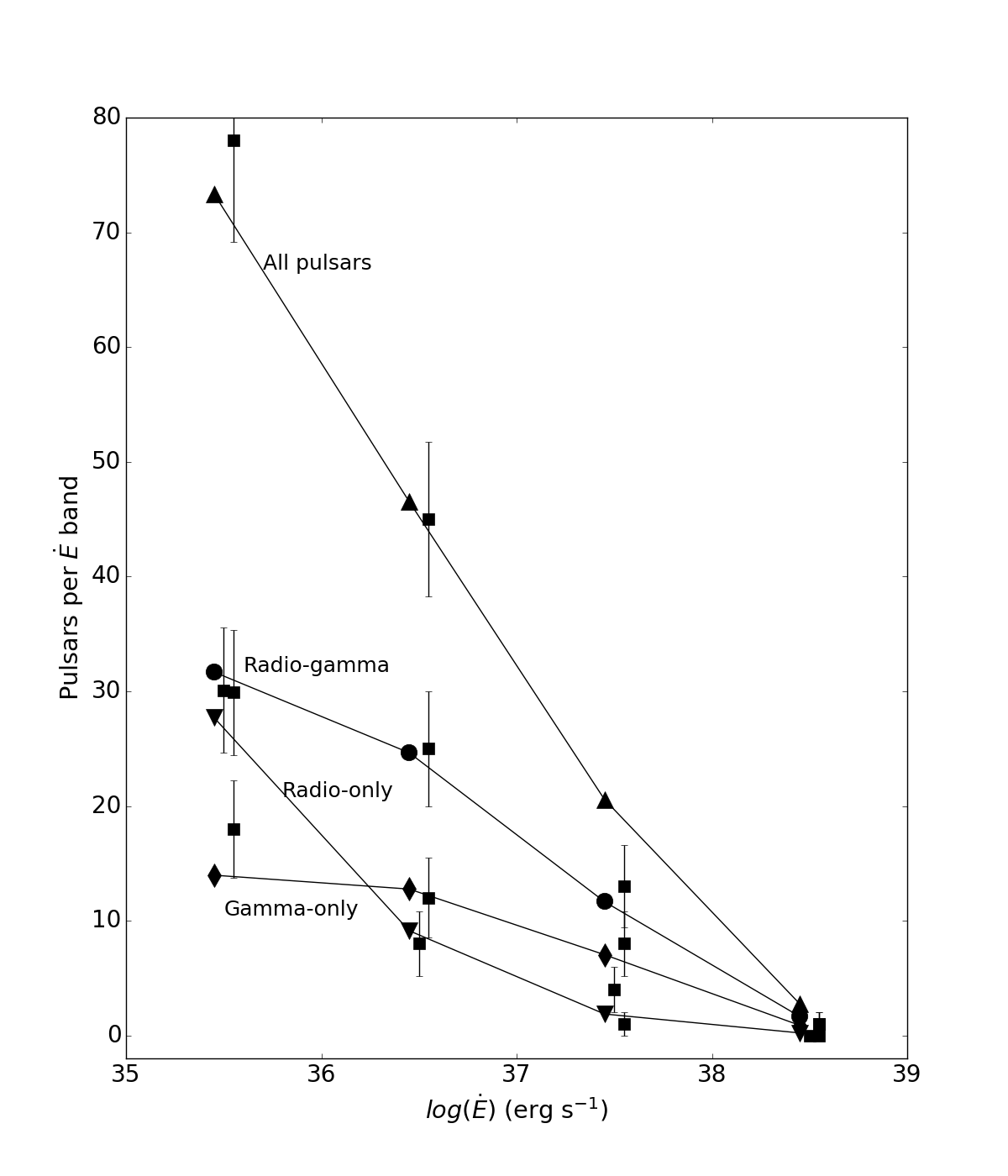}
\caption{A comparison between the observed number of pulsars and the numbers produced by the simulation for a birth rate of 1 per 95 years. Square symbols denote the observations (taken from Table~\ref{tab_edot} with $\sqrt{N}$ error bars. Downward triangles denote the radio-only pulsars, diamonds the $\gamma$-only pulsars, circles the radio/$\gamma$-ray sources and upward triangles the total numbers from the simulation (c.f Table~\ref{tab_bf2}).}
\label{fig:summary}
\end{figure}

\begin{figure}
\centering
\includegraphics[width=0.5\textwidth]{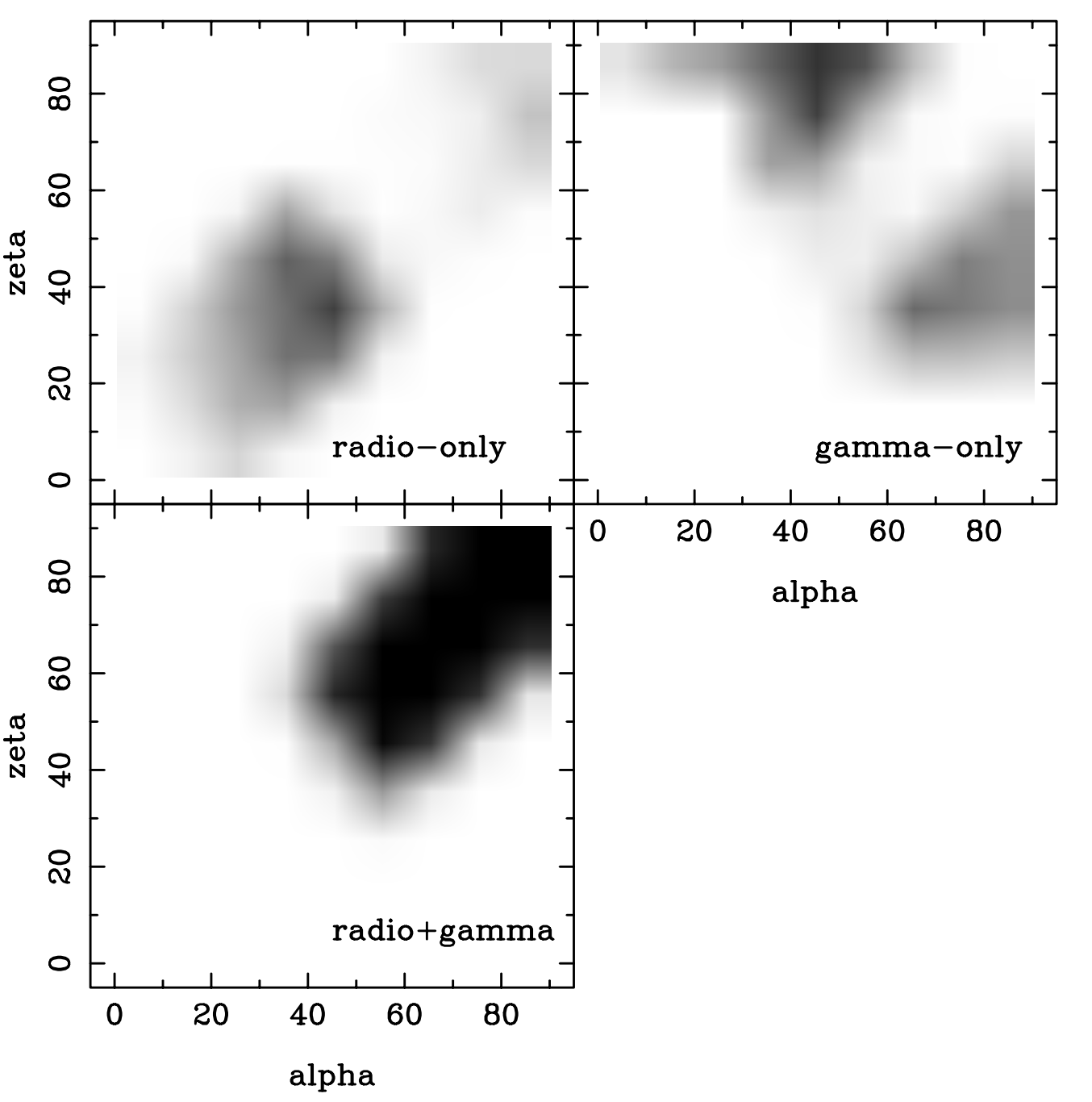}
\caption{Grey-scale representation of the detection of the various classes of pulsars in $\alpha$-$\zeta$ space for $\dot{E}=10^{36}$~erg s$^{-1}$. Darker shades represent higher detection numbers.}
\label{fig:gray}
\end{figure}

Figure~\ref{fig:summary} allows for a comparison of the number of pulsars in the various classes for the simulation and the observations. There is an excellent match for $\dot{E}<10^{37}$~erg s$^{-1}$. In particular, about 50\% of the pulsars are joint radio/$\gamma$-ray detections at $\dot{E}=10^{36}$~erg s$^{-1}$ with few radio-only pulsars, whereas at $\dot{E}=10^{35}$~erg s$^{-1}$ the numbers of radio-only and joint radio/$\gamma$-ray pulsars are similar. Figure~\ref{fig:gray} shows the geometry of the various classes of pulsars which are detected in the simulation in the $\dot{E}=10^{36}$ to $10^{37}$~erg s$^{-1}$ slice. The joint radio-$\gamma$-ray detections fall along the $\alpha=\zeta$ diagonal at high values of $\alpha$. The radio-only pulsars are seen at low $\alpha$ where $\gamma$-ray emission is not produced. The $\gamma$-ray only pulsars are seen away from the $\alpha=\zeta$ diagonal and have a preference for high $\zeta$.

\subsection{Where are the high $\dot{E}$ $\gamma$-ray pulsars?}
The main source of discrepancy between the simulation and the observations is that the simulation predicts a much higher fraction of $\gamma$-ray only pulsars and a lower fraction of radio only pulsars at $\dot{E}>10^{37}$~erg s$^{-1}$ than are seen in the observations. This was also remarked upon in the simulations of \citet{pghg12}. Could this be due to observational selection effects? We note that of the 4 radio-only pulsars, 3 of them have poor radio timing which means an optimum deep $\gamma$-ray search cannot be performed. There is also evidence that rotation irregularities in pulsars scale with the spin parameters and hence closely follow $\dot{E}$. Timing noise, which causes phase wander of the timing residuals, can be severe in the high $\dot{E}$ pulsars \citep{sc10,psj+19} which reduces the sensitivity in blind $\gamma$-ray searches. Furthermore, the occurrence of abrupt changes in spin period caused by glitches increases for higher values of $\dot{P}/P^2$ \citep{fer+17}. With the values of $\dot{E}$ and $P$ from Table~\ref{tab_sim} and the prescription from \citet{fer+17}, the estimated glitch rate of $\dot{E}\sim10^{37}$~erg s$^{-1}$ pulsars is 1 per 1.5~yr, whereas in $\dot{E}\sim10^{37}$~erg s$^{-1}$ pulsars this is 1 per 8.5~yr. Again this reduces sensitivity to blind $\gamma$-ray searches for higher $\dot{E}$ pulsars. The youngest pulsars may be located in supernova remnants or other regions of the Galactic plane with high background counts in $\gamma$-rays making it more difficult to detect the pulsed signal. Sources in these confused regions have larger position uncertainties, requiring greater computation effort to search. All these effect taken together provide at least a partial explanation for the lack of $\gamma$-ray only pulsars above $\dot{E}$ of $10^{37}$~erg s$^{-1}$ and although ways of mitigating these effects exist \citep{krj+15,cwp+17} it remains curious that the only $\gamma$-ray only pulsar at this $\dot{E}$ was found in the very early days of the {\it Fermi} mission \citep{spd+10}.

\subsection{Galactic birth rate}
The overall estimate of the Galactic birth rate of these fast-spinning neutron stars is 1 per $95\pm10$ year. The rate of core-collapse supernovae is estimated to be $1.9\pm1.1$ per century \citep{dhk+06}. Taking our result at face value means there is room to accommodate the creation of a further 1-2 neutron stars per century from core-collapse supernovae. These could manifest themselves as rotating radio transients (RRATs; \citealt{mll+06}), X-ray dim neutron stars (XDINs; \citealt{gh13}) or pulsars with long initial spin periods \citep{fk06}. This issue is comprehensively discussed in \citet{kk08}.

We also recall from Section~2 that there are pulsars seen in X-ray but not in the radio nor in $\gamma$-rays \citep{kh15}. There are two possible explanations. First, the pulsars were discovered in very deep X-ray exposures, they are in the Galactic plane and are distant. They could therefore potentially be $\gamma$-ray pulsars but lie below the Fermi LAT sensitivity. In this case, which we consider the most likely, our model accounts for them. The second possibility is that they lie in $\alpha,\zeta$ space where $\gamma$-rays are not expected (i.e. in the blank spaces in Figure~\ref{fig:gray}). From the simulations, the upper limit on such a population would be 25\% of the total. The birth rate could thus be greater by this amount.

\subsection{Radio and $\gamma$-ray beaming}
\citet{rmh10} compared the radio beaming fraction relative to the $\gamma$-ray beaming fraction via
\begin{equation}
R = \frac{N_{gr}}{N_g + N_{gr}}
\end{equation}
and their paper had $R=0.57\pm0.08$. At that time, $\gamma$-ray models implied that the $\gamma$-ray beaming fraction was expected to be of order unity, and \citet{rmh10} therefore postulated that the radio beaming fraction must be as high as 60\% for these type of pulsars. This implied high radio emission heights and/or the possibility of fan-beam emission rather than simple conal emission. Using the numbers in Table~\ref{tab_edot}, the modern value is $R=0.67\pm0.04$, consistent within the error bars with the earlier result. However, for the $\gamma$-ray model that we have used, the beaming fraction is much less than unity, particularly in the lowest $\dot{E}$ bin (see Figure~\ref{fig:CgZogPlot}). Table~\ref{tab_bf2} indeed shows that ratio $f_r/f_g$ is indeed above 0.60 at all $\dot{E}$, but the implication now is that the overall radio beaming fraction does not have to be as high as 60\% across the board.

\subsection{Radio emission heights}
The radio beaming is directly related to the corresponding emission heights (see Eqn.~\ref{eqn:rho}).
Hence, one striking result of this work is that the radio emission height can be kept fixed at 300~km for these young pulsars, a value similar to that found in the older population \citep{mr02}. However, it is important to note that relative to the light cylinder radius ($R_{\rm lc}$), this height is 1\% of $R_{\rm lc}$ in the old population but 6\% and 12\% of $R_{\rm lc}$ for a pulsar with a spin period of 100~ms or 50~ms respectively. These high fractional heights should manifest themselves in the observational data. For example, the effects of retardation and aberration (which are height dependent) mean that the inflexion point of the swing of the position angle of the linear polarization and the profile midpoint should be offset \citep{bcw91}. The magnitude of this offset, $\delta\phi({\rm PA})$, is given by
\begin{equation}
\delta\phi({\rm PA}) = \frac{4\,\,\, h_{\rm em}}{R_{lc}}
\end{equation}
and so for $h_{em}/R_{lc}=0.1$ this is $\sim$20\degr\ as opposed to only $\sim$2\degr\ for $h_{em}/R_{lc}=0.01$. \citet{jw06} used this technique to show that for some of these high $\dot{E}$ pulsars the heights did indeed exceed 5\% of $R_{\rm lc}$. However, the determination of the profile midpoint is not simple and \citet{wj08b} showed that emission heights computed via various methods showed no signs of agreement. However, at these high fractional heights, effects of magnetic sweepback also become important but difficult to deal with. The work of \citet{cr12} showed that radio emission heights computed conventionally would typically underestimate the true height.

\subsection{Radio interpulses}
The radio beaming and the pulsars' geometry also affects how many radio pulsars should show emission from both poles (i.e have interpulses). For instance, for a fixed emission height, pulsars with smaller spin periods have larger emission cones making it more likely that interpulses can be detected. In the observational set of 106 radio pulsars, only 6 show interpulse emission. The number expected from our simulation is 10.6, nearly a factor of two higher than the observations. Of the six, five are also seen in $\gamma$-rays with the exception being the binary pulsar PSR~J1906+0746 \citep{dkl+19}.  

We should perhaps not read too much into small number statistics, but this discrepancy is puzzling nonetheless. We note that the ratio of the amplitudes of the main and inter-pulses of radio pulsars can often be a factor of 10:1 or greater. Many of the joint radio and $\gamma$-ray pulsars are extremely weak radio sources and if the amplitude ratios were 3:1 or more then the interpulse may not have been detected. Deep radio observations of these weak sources are warranted.

Discrepancies may also be reflected by the aforementioned contrasting views as to whether $\alpha$ is varying with time and, if so, in which direction. Other, more fundamental explanations may be found in the physics relating to the generation of plasma that creates the radio emission. For instance, \cite{nbg+20} have recently suggested that the location of plasma-generating regions above the polar cap depends sensitively on $P$, magnetic field strength and, in particular, $\alpha$. The consequence would be, in their model, that both poles may not be associated with equally strong emission, thereby creating fewer observable interpulse pulsars than otherwise expected.  While this is potentially an interesting probe to understand the underlying emission physics, the interpulse statistics do not have an overall effect on our interpretation and conclusions. 

\subsection{Pulsars detected only in radio}
Figure~\ref{fig:gray} shows that (statistically) there is a difference in geometry between the joint $\gamma$-ray and radio detections and the radio-only detections. Radio pulsars at low $\alpha$ have, in general, wider profiles than those at high $\alpha$ as a consequence of Equation~\ref{eqn:wr}. In the simulations we do indeed see a strong demarcation between the widths of the radio profiles when comparing these two classes of detections.
This demarcation in radio profile widths above which $\gamma$-ray pulsars are not (or rarely) seen can be expressed as:
\begin{equation}
    W = 16.0\,\,\, {\rm log}(\frac{\dot{E}}{10^{35}}) + 40.0
\label{eq:rook}
\end{equation}
with W in degrees. In the observed population, 4 radio-only pulsars lie above this line, none of the joint radio/$\gamma$-ray pulsars do so. This was also shown in  figure~2 of \citet{rwjk17}. Their ``by-eye'' separation of the observed population of radio and radio/$\gamma$-ray pulsars has a somewhat steeper slope than equation~\ref{eq:rook}.

\subsection{Pulsars detected only in $\gamma$-rays.}
One point to note from the simulations is that once a "deep" radio survey is conducted on the $\gamma$-ray loud population, that very few potentially detectable radio pulsars remain undetectable. This is largely because the sensitivity of the radio searches far exceed those of the $\gamma$-ray surveys. This effectively means that the $\gamma$-ray only pulsars have high values of $|\zeta-\alpha|$, hence are not detectable as radio pulsars because the line of sight misses the (smaller) radio emission cone (i.e. that $|\zeta-\alpha| > \rho $). Figure~\ref{fig:gray} shows this effect clearly.

In Section~4.3 we described removing $\gamma$-ray pulsars from the simulation which only showed DC (and therefore not pulsed) emission. These DC sources will manifest themselves in the {\it Fermi} LAT point source catalogue but will not show up in pulsed searches. None of these sources are beamed towards Earth in the radio and there are no high $\dot{E}$ pulsars with only DC $\gamma$-ray counterparts. We find that some 18 DC sources are created in the simulation. Recently, \citet{wcp+18} searched over 100 $\gamma$-ray point sources, mostly at low latitudes, looking for pulsed signals. Although they had good success there remained some 80 sources unidentified and they surmise that some fraction of these must indeed be pulsars which produce only DC emission.

We note that the distribution of $\alpha$ in the $\gamma$-ray pulsars is more skewed towards orthogonal rotators as $\dot{E}$ decreases (Figure~\ref{fig:CgZogPlot}). Although not modelled in our simulation, the population of joint radio/$\gamma$-ray pulsars below $\dot{E}<10^{35}$~erg s$^{-1}$ should show this effect.
Indeed, \citet{hsg+14} came to a similar realisation in their successful attempt to find faint $\gamma$-ray pulsars with a small pulsed fraction at low $\dot{E}$. Determining the geometry through fits to the radio polarization of this lower $\dot{E}$ sample would make an interesting test of this prediction. 

\subsection{The GeV excess}
The Milky Way is very bright in GeV $\gamma$-rays. Cosmic rays incident on the gas and dust of the interstellar medium generate $\pi$ mesons, which then produce $\gamma$-rays by various channels. Around the Galactic centre, the diffuse emission observed by the {\it Fermi} LAT exceeds predictions by roughly 10\%. An abundant literature pits those who interpret this ``GeV excess'' as the signature of self-annihilating dark matter particles \citep[a recent example is][]{darkmatter} against those who argue that presently unknown pulsars could also generate the signal.

\citet{pulsarGeVexcess}, for example, model the contributions of both young and recycled pulsars (MSPs) to the diffuse emission. MSPs are less abundant and dimmer in gamma rays, are located at all Galactic latitudes, and shine essentially forever after recycling. Young pulsars shine only briefly in comparison, but are significantly brighter than MSPs and mostly reside within $2^\circ$ of the Galactic plane. The combined populations seem able to yield spatial and spectral gamma-ray distributions consistent with the observed GeV excess. Our simulation does not explicitly include a Galactic centre population but predicts a substantial number of high $\dot{E}$ pulsars, nearly all of which would be beaming $\gamma$-rays at Earth. Here we simply remark that this may mean a still larger contribution of pulsars to the diffuse emission, with a particularly hard gamma-ray spectrum, as per Figure 3 of \citet{pulsarGeVexcess}.

\subsection{Other models for radio and $\gamma$-ray emission and future work}
We have tested the conal model for radio emission and the outer-gap model for $\gamma$-ray emission. Other models exist. In the radio, the fan-beam model, first proposed by \citet{mic87} and taken up more recently by \citet{wpz+14} and \citet{dr15} is a viable alternative. In fast-spinning pulsars, the fan-beam provides a natural explanation for the high beaming fractions required without the need for the emission to arise from high in the magnetosphere.

In the $\gamma$-rays, promising progress has been made in the force-free models and pulsar profiles have been produced which match well with the observations \citep{ps18,kbt+18}. In broad terms, the FF model produces sky maps similar to the outer-gap model but with some differences. Similarities include that at low values of $\alpha$, emission is produced from the rotational equator, the beaming fraction is low and joint radio/$\gamma$-ray pulsars are not expected. However, for values of $\alpha$ near 90\degr, the beaming fraction is smaller in the FF models than for OG models.

One area of difference between the conal/fan-beam models in the radio and the OG/FF/TPC models in $\gamma$-rays is that they produce different pulsar profiles for the same geometry. In this paper we have not attempted to produce profiles and to compare them with observations as this requires detailed numerical simulations to cover the entire ($\alpha$,$\zeta$) plane. This is an obvious extension to the work carried out here.

We also note that the observed numbers of rotation-powered X-ray pulsars has increased in recent years and these provide yet another window into the high $\dot{E}$ population, especially as some are not seen at either $\gamma$-ray or radio wavelengths (e.g. \citealt{gh09}). Large-scale surveys of the sky in X-ray are very shallow, with most pulsars discovered in long-duration, targetted pointings of e.g. supernova remnants. This makes it hard to quantify selection effects. Recent observations of rotation powered pulsars by \citet{gkr+19} show phase-offsets between radio and X-ray profiles albeit in the millisecond pulsars. The location, luminosity and beaming of X-ray emission remains a fruitful avenue for future research.

Finally we have not attempted to discuss the population of millisecond pulsars even though these are a substantial fraction of the $\gamma$-ray sources. In the millisecond pulsars, the entire light cylinder is compressed into a few hundred km and a large fraction of the field lines are open. In spite of the differences in spin period and magnetic field strength, the millisecond pulsars mimic the behaviour of their slower spinning counterparts. Even so, our understanding of the shape of the radio beam, its filling factor and polarization properties is not at all well constrained, making modelling difficult.

\section{Summary}
We have used the {\it Fermi} survey of pulsars in conjunction with the radio surveys of the Galactic plane to understand the population of young, high $\dot{E}$ pulsars. Somewhat to our surprise, we found that simple models of the radio and $\gamma$-ray emission and some basic assumptions about the underlying population are sufficient to reproduce the observed results. We estimate that there are some 850 pulsars in the Galaxy with $\dot{E}>10^{35}$~erg s$^{-1}$ and that the birth rate of fast-spinning pulsars is 1 per 95 years in line with the core-collapse supernova remnant rate.

There appears to be fewer $\gamma$-ray pulsars observed at $\dot{E}>10^{37}$~erg s$^{-1}$ than predicted. We surmise this is due to observational selection effects, primarily because of the increased glitch rate in these pulsars. There also appears to be fewer radio pulsars showing interpulse emission than expected and we discuss a possible reason to do with the emission physics.

We find that the beaming fraction of radio pulsars is 0.6 that of the beaming fraction in $\gamma$-rays. This was asserted by \citet{rmh10}, but they made the incorrect assumption that the $\gamma$-ray beaming fraction was of order unity. We therefore do not need to appeal to fan-beam models to explain the data. Instead we find that an emission height of 300~km can be used for these pulsars, similar to the radio pulsar population as a whole. This height is a substantial fraction of the light cylinder radius for pulsars with spin periods less than 100~ms. The outer-gap model provides a good fit to the $\gamma$-ray data, and our results are broadly in line with those of \citet{wr11} except that our birth rate is somewhat lower.

In future, a detailed comparison of different emission models for radio and $\gamma$-rays, particularly with respect to generating pulse profiles and by including the population of millisecond pulsars is warranted. Progress with modelling of rotation powered X-ray pulsars will also be fruitful. As far as increasing the observed population is concerned, {\it Fermi} will continue to slowly uncover more $\gamma$-ray pulsars as more compute power is thrown at blind searches. In the radio, deeper surveys of the Galactic plane are planned with the Parkes and MeerKAT telescopes in the lead up to the major survey with the Square Kilometre Array.

\section*{Acknowledgements}
We thank Colin Clark for fruitful discussions and the referee for useful comments.

\section*{Data Availability}
The data underlying this article will be shared on reasonable request to the corresponding author.



\bibliographystyle{mnras}
\bibliography{gray_edot} 


\bsp	
\label{lastpage}
\end{document}